\documentclass[12pt]{iopart}
\usepackage{graphicx}
\usepackage{iopams}
 
\begin{document}

\title{Expansion of the Universe - Standard Big Bang Model}

\author{Matts Roos}

\address{Department of Physical Sciences and Department of
          Astronomy\\ FIN-00014 University of Helsinki, Finland}
\ead{matts.roos@helsinki.fi}
\begin{abstract}After a brief introduction to the sixteenth and seventeenth century views of the Universe and the nineteenth century paradox of Olbers, we start the history of the cosmic expansion with Hubble's epochal discovery of the recession velocities of spiral galaxies. By then Einstein's theories of relativity were well known, but no suitable metric was known. Prior to introducing General Relativity we embark on a non-chronological derivation of the Robertson-Walker metric directly from Special Relativity and the Minkowski metric endowed with a Gaussian curvature. This permits the definition of all relativistic distance measures needed in observational astronomy. Only thereafter do we come to General Relativity, and describe some of its consequences: gravitational lensing, black holes, various tests, and the cornerstone of the standard Big Bang model, the Friedmann-Lema\^{\i}tre equations. Going backwards in time towards Big Bang we first have to trace the thermal history, and then understand the needs for a cosmic inflation and its predictions. The knowledge of the Big Bang model is based notably on observations of the Cosmic Microwave Background Radiation, large scale structures, and the redshifts of distant supernovae. They tell us that gravitating matter is dominated by a dark and dissipationless component of unknown composition, and that the observable part of the Universe exhibits an accelerated expansion representing a fraction of the energy even larger than gravitating matter.
\end{abstract}
\maketitle
\section*{Contents}
1. Historical cosmology\\
2. Olbers' paradox\\
3. Hubble's law\\
4. Special Relativity and metrics\\
5. Distance measures\\
6. General Relativity\\
7. Tests of General Relativity\\
8. Gravitational lensing\\
9. Black holes\\
10. Friedmann-Lema\^{\i}tre cosmology\\
11. Thermal history of the Universe\\
12. Cosmic inflation\\
13. Cosmic Microwave Background Radiation\\
14. Large Scale Structure\\
15. Dark Matter\\
16. Dark Energy\\
17. Conclusions

 \section{Historical cosmology}
The history of ideas on the structure and origin of the Universe shows that humankind has always put itself at the center of creation. As astronomical evidence has accumulated, these \emph{anthropocentric} convictions have had to be
abandoned one by one. From the natural idea that the solid Earth is at rest and the celestial objects all rotate around
us, we have come to understand that we inhabit an average-sized planet orbiting an average-sized sun, that the Solar
System is in the periphery of The Milky Way, a rotating galaxy of average size, flying at hundreds of kilometers
per second towards an unknown goal in an immense Universe, containing billions of similar galaxies.

Cosmology aims to explain the origin and evolution of the entire contents of the Universe, the underlying physical
processes, and thereby to obtain a deeper understanding of the laws of physics assumed to hold throughout the Universe.
Unfortunately, we have only one universe to study, the one we live in, and we cannot make experiments with it, only
observations. This puts serious limits on what we can learn about the origin. If there were other universes we would never know.

Although the history of cosmology is long and fascinating we shall not trace it in detail, nor any further back than \textit{Isaac Newton} (1642--1727).
In the early days of cosmology when little was
known about the Universe, the field was really just a branch of philosophy.
At the time of Newton the heliocentric Universe of
\textit{Nicolaus Copernicus} (1473--1543), \textit{Galileo Galilei}
(1564--1642) and \textit{Johannes Kepler} (1571--1630) had been accepted because no sensible
description of the motion of the planets could be found if the Earth was at rest at the center of the Solar
System. However, this anthropocentric view persisted, locating the Solar
System at the center of the Universe.
The Milky Way had been resolved into an accumulation of faint stars with the telescope of
Galileo. Copernicus had formulated the \textit{cosmological} or \textit{Copernican principle}, according to which
\begin{itemize}\item\textit{the Universe is homogeneous and isotropic in three-dimensional space, has always been so, and will always remain
so.}
\end{itemize}
Obviously, matter introduces lumpiness which violates homogeneity on the scale of stars and on the scale of the Milky Way, but on
some larger scale isotropy and homogeneity is still taken to be a good approximation.

The first theory of gravitation appeared when Newton published his \textit{Philosophiae Naturalis Principia
Mathematica} in~1687, explaining the empirical laws of Kepler: that the planets moved in
elliptical orbits with the Sun at one of the focal points. Newton considered the stars to be suns like ours, evenly distributed in a static, infinite Universe. The total number of stars could not be infinite
because then their attraction would also be infinite, making the static Universe unstable. There were controversial opinions whether the number of stars was finite or infinite, and whether a finite universe was bounded and an infinite one unbounded.
Later \textit{Immanuel Kant} (1724--1804) claimed that the question of infinity was irrelevant because neither type of system embedded in infinite space could be stable and
homogeneous. The right conclusion is that the Universe cannot be static, an idea which would have been too revolutionary at Newton's time.
The infinity argument was, however, not properly understood until \textit{Bernhard Riemann} (1826--1866) pointed out that the world could be \textit{finite} yet \textit{unbounded}, provided the geometry of the space had a positive curvature, however small.

The first description of the Milky Way as a rotating galaxy can be
traced to \textit{Thomas Wright} (1711--1786). Wright's galactic picture had a direct impact on Kant who suggested in 1755 that the diffuse nebulae observed by Galileo could be distant
galaxies rather than nearby clouds of incandescent gas. This implied that the Universe could indeed be homogeneous on the
scale of galactic distances. This view was also defended by \textit{Johann Heinrich
Lambert} (1728--1777) who came to the conclusion that the Solar System, along with the
other stars in our Galaxy, orbited around the galactic center, thus departing from the heliocentric view. Kant and Lambert thought that matter is clustered on ever larger scales of hierarchy and that matter is endlessly being recycled.
This leads to the question of the origin of time: what was the first cause of the rotation of the galaxy and when did it all start? This is the question modern cosmology attempts to answer by tracing the evolution of the Universe backwards in time.

Newton's first law states that \textit{inertial systems} on which no forces act are either at rest or in uniform motion. He considered that these properties implicitly referred to an absolute space that was unobservable, yet had a real existence. In 1883 \textit{Ernst Mach} (1838--1916) rejected the concept of absolute space, precisely because it was
unobservable: the laws of physics should be based only on concepts which could be related to
observations. Since motion still had to be referred to some frame at rest, he proposed replacing absolute space by an idealized rigid frame of fixed stars. Although Mach clearly realized that all motion is relative, it was left to \textit{Albert
Einstein} (1879--1955) to take the full step of studying the laws of physics as seen by observers in inertial frames in relative motion with respect to each other. On the basis of Riemann's geometry, Einstein subsequently established the connection between the geometry of
space and the distribution of matter.

In spite of the work of Kant and Lambert, the heliocentric picture of the Galaxy remained well into the 20th century. A decisive change came with the
observations in 1915--1919 by \textit{Harlow Shapley} (1895--1972) of the distribution of
\textit{globular clusters} hosting $10^5$--$10^7$~stars. He found that perpendicular to the
galactic plane they were uniformly distributed, but along the plane these clusters had a distribution which peaked in the direction of the Sagittarius. This defined the center of the Galaxy to be quite far from the Solar System: we are at a distance of about two-thirds of the galactic radius. Thus the anthropocentric world picture received yet another
blow, and not the last one. Shapley still
believed our Galaxy to be at the center of the astronomical Universe.

\section{Olbers' paradox}
An early problem still discussed today is the paradox of \textit{Wilhelm
Olbers} (1758--1840): why is the night sky dark if the
Universe is infinite, static and uniformly filled with stars? They should fill up the total field of visibility so that the night sky would be as bright as the Sun, and we would find ourselves in the middle of a heat bath of the temperature of the surface of the Sun. Obviously, at least one assumption about the Universe must be wrong.

Olbers' own explanation was that invisible interstellar dust absorbed the starlight so as to make its intensity decrease exponentially with distance. But one can show that the amount of dust needed would be so great that the Sun would also be obscured. Moreover, radiation heats dust so that it becomes visible in the infrared.

A large number of different solutions to this paradox have been proposed, and indeed several effects can be invoked (see ref. Harrison). One possible explanation evokes expansion and special relativity. If the Universe expands, starlight redshifts, so that each arriving photon carries less energy than when it was emitted. At the same time, the volume of the Universe grows, and thus the energy density decreases. The observation of the low level of radiation in the intergalactic space has in fact been evoked as a proof of the expansion.

The dominant effect is, however, that stars radiate only for a finite time, they burn their fuel at well-understood rates. Each galaxy has existed only for a finite time, whether the age of the Universe is infinite or not. Also, the volume of the observable Universe is not infinite, it is in fact too small to contain
sufficiently many visible stars. When the time perspective grows, an
increasing number of stars becomes visible because their light has had time to reach us, but at the same time stars which have burned their fuel disappear.

\section{Hubble's law}
In a static universe the galaxies should move about randomly, but early galaxy
observations had shown that atomic spectral lines of known wavelengths $\lambda$ exhibited a systematic redward shift to $\lambda '$ by a factor $1+z=\lambda '/\lambda$ (an exception is the blueshifted \emph{Andromeda nebula}~M\,31), thus these galaxies were receding from us with velocity $v=cz$. In an expanding homogeneous Universe distant galaxies should appear to recede faster than nearby ones.

In the 1920\,s  \textit{Edwin P. Hubble} measured the recession velocities of 18~spiral galaxies with a reasonably well-known distance, and found that all
the velocities increased linearly with distance, $v=H_0r$, or
\begin{equation}
z=H_0\frac{r}{c}.
\end{equation}
This is \textit{Hubble's law}, and $H_0$ is called the \textit{Hubble
parameter} (present values are always subscripted 0). The message of Hubble's law is that the Universe is expanding and a static Universe is thus ruled out. Einstein had until then firmly believed in a static universe, but when he met Hubble in 1929 he was
overwhelmed. This moment marks the beginning of modern cosmology, and sets the primary requirement on theory.

The expansion affects the wavelengths of radiation and the distances between galaxies, but it does not affect the size and internal distances of gravitationally bound systems such as the Solar system, the Milky Way or other galaxies. The expansion appears as if all astronomical objects were receding from us and we were at the center of the Universe. But the Cosmological Principle does not allow a center, and therefore every observer, regardless of position, will have the same impression. Thus the observed recession is really a general expansion.

Equation~(1) shows that the Hubble parameter has the dimension of inverse time. Thus
a characteristic timescale for the expansion of the Universe is the \textit{Hubble time} $\tau_{\rm H}=H_0^{-1}$, and the size scale of the observable Universe is the \textit{Hubble
radius} $r_{\rm H}=\tau_{\rm H}\, c$. In Section 5 we shall discuss measurements of $H_0$.
Using the dimensionless quantity $h=H_0/(100{\rm \, km~s^{-1}Mpc^{-1}})$ which has the value $h\approx 0.72$, we can derive
\begin{equation}
\tau_{\rm H}\equiv H_0^{-1}=9.78\, h^{-1}\times10^{9}{\rm yr},~~~~r_{\rm H}\equiv\tau_{\rm H}\, c=3000h^{-1}{\rm Mpc}.
\end{equation}
Radiation travelling with the speed of light $c$ reaches $r_{\rm{H}}$ in time $\tau_{\rm{H}}$. Note that Hubble's law is non-relativistic, objects beyond  $r_{\rm H}$ would be expected to
attain recession velocities exceeding $c$, which is an absolute limit in the theory of special relativity.

The size of the expanding Universe is unknown and unmeasurable, but it is convenient to express distances at different epochs in terms of a
\textit{cosmic scale factor}: at time $t$ the scale was $a(t)$ when the present value is $a(t_0)\equiv a_0\equiv 1$. The rate of change of the scale factor can then be identified with the Hubble parameter, $H(t)=\dot{a}(t)/a(t)$ (to first-order time differences).

\section{Special Relativity and metrics}
In Einstein's theory of special relativity one studies how signals are exchanged between inertial frames in motion with constant velocity with
respect to each other. Einstein postulated that
\begin{itemize}
\item
the results of measurements in different frames must be identical, and
\item
light travels by a constant speed in vacuo, $c$, in all frames.
\end{itemize}

Consider two linear axes $x$ and $x'$ in one-dimensional space, $x'$ being at rest and $x$ moving with constant
velocity $v$ in the positive $x'$~direction. Time increments are measured in the two coordinate systems as ${\rm d}t$ and
${\rm d}t'$ using two identical clocks. Neither the spatial increments ${\rm d}x$ and
${\rm d}x'$ nor the time increments are invariants -- they do not obey the first postulate. Let us replace ${\rm d}t$ and ${\rm d}t'$  with the \emph{temporal distances} $c\,{\rm d}t$ and $c\,{\rm d}t'$  and look for a linear transformation between
the primed and unprimed frames under which the two-dimensional \textit{spacetime distance} element ${\rm d}\bf{s}$ between two \textit{spacetime events},
\begin{equation}
{\rm d}s^2=c^2\,{\rm d}\tau^2\equiv c^2\,{\rm d}t^2-{\rm d}x^2=c^2\,{\rm d}t^{\prime 2}-{\rm d}x^{\prime 2}\equiv c^2\,{\rm d}\tau^{\prime 2},
\end{equation}
is invariant. The quantity ${\rm d}\tau$ is called the \textit{proper time} and ${\rm d}\bf{s}$
the \textit{line element}.

Invoking the second postulate it is easy to show that the transformation must be of the form
\begin{equation}
{\rm d}x'=\gamma\,({\rm d}x-v\,{\rm d}t),~~~~ c\,{\rm d}t'=\gamma\,(c\,{\rm d}t-v\,{\rm d}x/c),
\end{equation}
where
\begin{equation}
\gamma=(1-(v /c)^2)^{-1/2}.
\end{equation}

Equation~(4) defines the \textit{Lorentz transformation}, after \textit{Hendrik
Antoon Lorentz} (1853--1928). Scalar pro\-ducts (such as ${\rm d}\tau^2$ and ${\rm d}x^2$) in this two-dimensional
$(ct,x)$-spacetime are invariants under Lorentz transformations. For example, a particle with mass $m$ moving with velocity three-vector ${\bf v}$ and three-momentum ${\bf p}=m{\bf v}$ is described in four-dimensional spacetime by the four-vector $P=(E/c,~{\bf p})$. The scalar product $P^2$ is an invariant related to the mass,
$P^2=(E/c)^2-p^2=(\gamma mc)^2$. For a particle at rest, this gives Einstein's famous formula \begin{equation}
E=mc^2.
\end{equation}

It follows that time intervals measured in the two frames are related by ${\rm d}t=\gamma\,{\rm d}t'$.  This \textit{time dilation effect} is only noticeably when $v$
approaches $c$. It has been confirmed in particle accelerators and by muons produced in cosmic ray collisions in the upper atmosphere. These unstable particles have well-known lifetimes in the laboratory, but when they strike Earth with relativistic velocities, they appear to have a longer lifetime by
the factor~$\gamma$.

The Lorentz transformations (4) can immediately be generalized to three spatial coordinates $x$, $y$, $z$, so that the \textit{metric} (3) is replaced by the four-dimensional metric of
\textit{Hermann Minkowski} (1864--1909),
\begin{equation}
{\rm d} s^2=c^2\,{\rm d}\tau^2=c^2\,{\rm d}t^2-{\rm d}x^2-{\rm d}y^2-{\rm d}z^2\equiv c^2\,{\rm d} t^2-{\rm d} l^2.
\end{equation}

The trajectory of a body moving in spacetime is called its \textit{world line}. A body at a fixed location in space follows a world line parallel to the time axis in the direction of increasing time. A moving
body follows a world line making a slope with respect to the time axis. Since the speed of a body or a
signal travelling from one event to another cannot exceed the speed of light, there is a maximum slope to such world
lines. All world lines for which $ct<0$ and arriving at $t=0$ form our \textit{past light
cone}, thus they enclose the present observable
universe. All world lines for which $ct>0$ and starting from where we are now can influence events inside our \textit{future light
cone}. Two separate events in spacetime can be causally connected provided their spatial separation ${\rm d}\bf{l}$ and
their temporal separation ${\rm d}t$ (in any frame) obey $|{\rm d}\bf{l}/{\rm d}t|\leq c.$
Their world line is then inside the light cone.
In Figure~1 we draw this four-dimensional cone in $t,x,y$-space (suppressing the $z$ direction).
  \begin{figure}[htbp]
\includegraphics{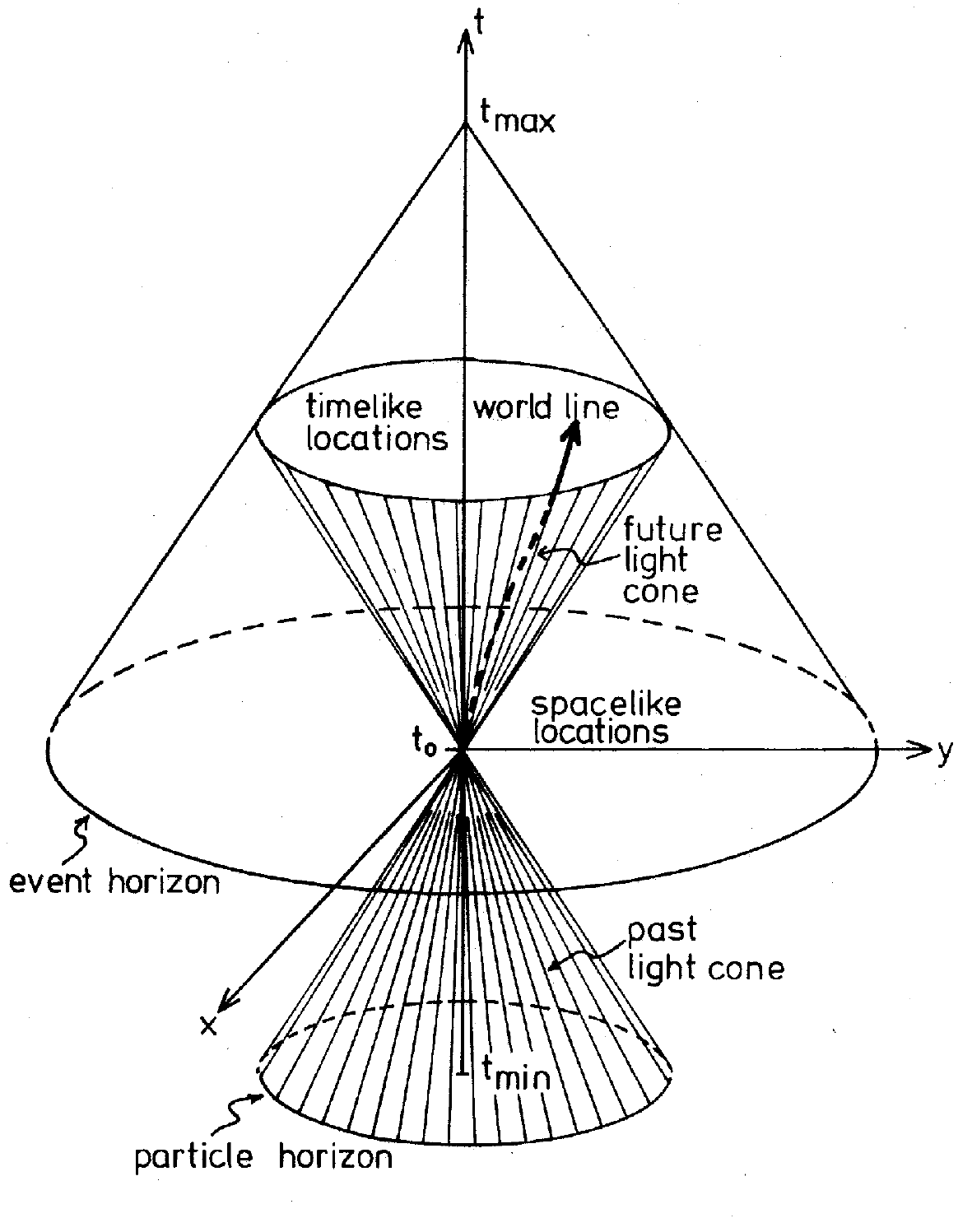}
\caption{Light cone in $x,y,t$-space. An event which is at the origin $x=y=0$ at the present time $t_0$ will follow some world line into the future, always remaining inside the future light cone. All points on the world line are at time-like locations with respect to the spatial origin at $t_0$. World lines for light signals emitted from (received at) the origin at $t_0$ will propagate on the envelope of the future (past) light cone. No signals can be sent to or received from space-like locations.
The space in the past from which signals can be received at the origin is restricted by the particle horizon, located at the earliest time under consideration. The event horizon restricts the space which can at present be in causal relation to the present spatial origin at some future time $t_{\rm max}$.}
    \end{figure}

Special relativity thus revised our concept of spacetime and made it four-dimensional. Riemann and others realized that
Euclidean geometry was just a particular choice suited to flat space, but not necessarily correct in the space we inhabit. Consider the path in three-space followed by a free body obeying Newton's first law of
motion. This path represents the shortest distance
between any two points along it, called a \textit{geodesic} of the space.
In flat Euclidean space the geodesics are straight lines. But measurements of distances depend on the geometric properties of space, as has been known to navigators ever since
Earth was understood to be spherical. A spherical surface is characterized by its radius of curvature which causes the geodesics to be great circles.

Suppose an observer wants to make a map of points in the expanding Universe. It is then no longer convenient
to use the coordinates $x,~y,~z$ in Equations~(3) and~(7) nor the spherical coordinates $R,~\theta,~\phi$, because the cosmic expansion would quickly outdate the map. Instead it is convenient to factor out the expansion $a(t)$ and replace the radial distance $R$ by $a(t)\sigma$, where $\sigma$ is a dimensionless stationary \textit{comoving} coordinate.

If the four-dimensional space happens to be curved just like the surface of Earth, a \emph{Gaussian curvature k} may be included in the Minkowski metric. The parameter $k$ can take on the values $+1$, $0$, $-1$, corresponding to a three-sphere, a flat three-space, and a three-hyperboloid, respectively. The metric of four-dimensional spacetime can then be written in the form derived independently by \textit{Howard Robertson} and \textit{Arthur Walker} in 1934:
\begin{equation}
{\rm d} s^2=c^2\,{\rm d} t^2-{\rm d} l^2=c^2\,{\rm d} t^2-a(t)^2\bigg(\frac{{\rm d}\sigma^2}{1-k\sigma^2}+\sigma^2{\rm d}\theta^2+
\sigma^2\sin^2\theta\,{\rm d}\phi^2\bigg).
\end{equation}
This metric (RW) can describe an expanding, spatially homogeneous and isotropic universe in accord with the cosmological principle.

Of course there was a motivation for introducing curvature: General Relativity, to which we shall come in Section 6.
\section{Distance measures}
The \textit{comoving
distance} from us to a galaxy at comoving coordinates $(\sigma,0,0)$ is not an observable
because the galaxy can only be observed by the light it emitted at an earlier time, $t<t_0$. In a spacetime described by the RW metric the light signal propagates along a geodesic, ${\rm d} s^2=0$. Introducing an alternative comoving coordinate $\chi$ defined by ${\rm d}\chi={\rm d}\sigma/\sqrt{1-k \sigma^2}$ in Equation~(8) (with ${\rm d}\theta^2={\rm d}\phi^2=0$), the geodesic equation is ${\rm d} s^2 = c^2{\rm d}t^2-a(t)^2{\rm d}\chi^2=0$. From this,
\begin{equation}
\chi=c\int_t^{t_0}\frac{{\rm d}t}{a(t)}.
\end{equation}

The present \textit{proper distance} to the galaxy is then $d_{\rm P}=a_0\chi$.
In flat space $d_{\rm P}=\sigma$, but in curved spaces the function is more complicated (cf. Roos). For practical astronomical measurements at small redshifts one uses an approximate expression for $d_{\rm P}$ in flat space ($k=0$),
\begin{equation}
d_{\rm P}(z)\approx {c\over H_0}\left(z-\frac 12 (1+q_{\,0})z^2\right),
\end{equation}
where $q_{\,0}$ is the present \textit{deceleration parameter}
$q\equiv -a\ddot{a}/\dot{a}^2=-\ddot{a}/aH^2$. The first term on the right of Equation~(10) gives Hubble's linear law (1), and thus the second term measures deviations from
linearity to lowest order. The parameter value $q_{\,0}=-1$ obviously corresponds to no deviation.

The largest comoving spatial distance  from which a signal could have reached us is called the \textit{particle horizon}, denoted  $\sigma_{\rm ph}$ or alternatively $\chi_{\rm ph}$. This delimits the part of the Universe that has come into causal contact since time $t=0$. If the lower integration limit in Equation~(9) is equal to the time when the Universe became transparent to light (the last scattering time), the particle horizon delimits the visible Universe.

In an analogous way, the comoving distance $\sigma_{\rm eh}$ or $\chi_{\rm eh}$ to the \textit{event horizon} is defined as the spatially most distant event at time $t_0$ from which a world line can ever reach our world line. By `ever' we
mean a finite future time, $t_{\rm max}$. The particle horizon $\sigma_{\rm ph}$ at time $t_0$ lies on our past light cone, but with time our particle horizon
will broaden so that the light cone at $t_0$ will move inside the light cone at $t>t_0$.
The event horizon at this moment can only be specified given the time distance to the ultimate future, $t_{\rm max}$.
Only at $t_{\rm max}$ will our past light cone encompass the present event horizon. Thus the event horizon is our ultimate particle horizon.

The distances to relatively nearby stars can
be measured by the \textit{trigonometric parallax} up to about 30~pc away. This is the difference in angular position of a star as seen
from Earth when at opposite points in its circumsolar orbit. The \textit{parallax distance} is defined as $d_{\rm par}=d_{\rm P}/\sqrt{1-k\sigma^2}.$

Consider an astronomical object radiating photons isotropically with power or absolute luminosity~$L$. At the \textit{luminosity distance} $d_{\rm L}$ from the object we observe only the fraction $B=L/4\pi d_{\rm L}^{\,2}$, its surface brightness, given by the Euclidean inverse-square distance law. If the Universe does not expand and the object is stationary at proper distance $d_{\rm P}$, a telescope
with area $A$ will receive a fraction $A/4\pi d_{\rm P}^2$ of the photons. But in a universe characterized by an
expansion $a(t)$, the object is not stationary, so the energy of photons emitted at time $t_{\rm e}$ is redshifted by
the factor $(1+z)=a^{-1}(t_{\rm e})$. Moreover, the arrival rate of the photons suffers time dilation by
another factor $(1+z)$, often called the \textit{energy effect}. The end result is that $d_{\rm L}=d_{\rm P}(1+z)$.

Astronomers usually replace $L$ and $B$ by two empirically defined quantities, \textit{absolute
magnitude} $M$ of a luminous object and \textit{apparent magnitude}
$m$. The replacement rule is
\begin{equation}
m-M=-5+5\log d_{\rm L},
\end{equation}
where $d_{\rm L}$ is expressed in parsecs (pc) and the logarithm is to base~10.

Most stars in the Galaxy for which we know $L$ from a kinematic distance determination exhibit a relationship between surface temperature $T$ and $L$, the
\textit{Hertzsprung--Russell} relation. These
\textit{main-sequence stars} sit on a fairly well-defined curve in the
$T-L$ plot, and temperature is related to color. From this relation one can derive
distances to farther main-sequence stars: from their color one obtains the luminosity which subsequently determines
$d_{\rm L}$. By this method one gets a second rung in a ladder of estimates which covers distances within our Galaxy.

Yet another measure of distance is the \textit{angular size distance} $d_{\rm A}$. In
Euclidean space an object of size $D$ which is at distance $d_{\rm A}$ will subtend an angle $\theta$ such that $\theta\approx D/d_{\rm A}$ for small angles.
In General Relativity we can still use this approximation to define $d_{\rm A}$. From the RW metric~(8) the diameter of a source of light at comoving distance $\sigma$ is $D=a\sigma\theta$, so $d_{\rm A}=D/\theta=a\sigma=\sigma/(1+z)$.

As the next step on the distance ladder one chooses calibrators which are stars or astronomical systems with specific
uniform properties, so called \textit{standard candles}. The \textit{RR~Lyrae}
stars all have similar absolute luminosities, and they are bright enough to be seen out to about 300~kpc. A very
important class of standard candles are the \textit{Cepheid} stars, whose absolute luminosity
oscillates with a constant period $\log P\propto 1.3\log L$.
Globular clusters\index{globular clusters} are gravitationally bound systems of $10^5$--$10^6$ stars forming a
spherical population orbiting the center of our Galaxy. They can also be seen in many other galaxies, and they are visible out to 100~Mpc. Various statistical properties of
well-measured clusters, such as the frequency of stars of a given luminosity, the mean luminosity, and the maximum luminosity are presumably shared by similar clusters at all
distances, so that clusters become standard candles. Similar statistical indicators can be used to calibrate clusters of galaxies; in particular the brightest
galaxy in a cluster is a standard candle useful out to 1~Gpc.

A notable contribution to our knowledge of $H_0$ comes from the observation of Type~Ia \textit{supernova} explosions. The released energy is always nearly the same, in particular
the peak brightness of Type~Ia supernovae can serve as remarkably precise standard candles out to~1~Gpc. Additional information is provided by the color, the
spectrum, and an empirical correlation observed between the timescale of the supernova light curve and the peak
luminosity.

The existence of different methods of calibration covering similar distances is a great help in achieving higher
precision. The expansion can be verified by measuring the surface brightness of standard candles at varying redshifts,
the \textit{Tolman test}. In an expanding universe, the intensity of the photon signal at
the detector is further reduced by a factor $(1+z)^2$ due to an optical aberration which makes the surface area of the
source appear increased. Such tests have been done and they do confirm the expansion.

The \textit{Tully--Fisher} relation is a very important tool at distances which overlap
those calibrated by Cepheids, globular clusters, galaxy clusters and several other methods. This empirical relation
expresses correlations between intrinsic properties of whole spiral galaxies. It is observed that their absolute
luminosity and their circular rotation velocity $v_{\rm c}$ are related by
$L\propto v_{\rm c}^4$. (For more details on this Section, see the books by
Peacock and Roos.)

The  differential light propagation delay between two or more gravitationally lensed images of a background object such as a quasar establishes an absolute physical distance scale in the lens system. This is named the \emph{Refsdal Method}, and it is the only direct way of measuring cosmological distances and the global expansion rate $H_0$ in a single step for each system, thus avoiding the propagation of errors along the distance ladder. There are about 10 such cases measured by now.

\section{General Relativity}
Although \textit{Newton's second law}, ${\bf F}=m{\bf a}$, is invariant under special relativity in any inertial frame, it is not invariant in accelerated frames because it explicitly involves acceleration, ${\bf a}$. Einstein required that also observers in accelerated frames should be able to agree on the value of acceleration. Space-time derivatives in a curved RW metric are also not invariants because they imply transporting quantities along some curve and that makes them coordinate~dependent. Thus the next
necessary step is to search for invariant redefinitions of derivatives and accelerations, and to formulate the laws of physics in terms of them. Such a formulation is called \textit{generally covariant}.
Moreover, for a body of \textit{gravitating mass} $m_{\rm G}$ at a distance $r$ from another mass $M$, the force $F$ specified by \textit{Newton's law of gravitation},
\begin{equation}
F=-GMm_{\rm G}/{r^2},
\end{equation}
where $G$ is \emph{Newton's constant}, is in serious conflict with special relativity in three
ways. Firstly, there is no obvious way of rewriting the law in terms of invariants, since it only contains scalars. Secondly, it has no explicit time dependence, so gravitational effects propagate instantaneously to every location in the Universe.
Thirdly, $m_{\rm G}$ is
totally independent of the \textit{inert mass $m$} appearing in Newton's second law, yet for unknown reasons both masses appear to be equal to a precision of $10^{-13}$ or better. Clearly a theory is needed to establish a formal link between them.

Einstein considered how Newton's laws would be understood by a passenger in a spacecraft, and realized that the passenger would not be able to distinguish between gravitational pull and local acceleration -- this is called the \textit{Weak Equivalence Principle} (WEP). This principle is already embodied in the \textit{Galilean equivalence principle} in mechanics between motion in a uniform gravitational field and a uniformly accelerated frame of reference. What Einstein did was to generalize this to all of physics, in particular phenomena involving light.
The more general formulation is the important \textit{strong equivalence principle } (SEP), that
\begin{quote}
\textit{to an observer in free fall in a gravitational field the results of all local experiments are completely independent of the magnitude of the field}.
\end{quote}

In a suitably small spacecraft, curved spacetime can always be locally approximated by flat Minkowski spacetime. On a larger scale a nonuniform
gravitational field can be replaced by a patchwork of locally flat frames which describe the curved space. Trajectories of bodies as well as rays of light follow geodesics, thus in a curved spacetime also light paths are curved. Following SEP, this implies that photons in a gravitational field may appear to have mass.

In the gravitational field of Earth, two test bodies
with a space-like separation clearly do not fall along parallels, but along different radii, so that their separation decreases with time. This phenomenon is called the \textit{tidal effect}, or
the tidal force, since the test bodies move as if an attractive exchange force acted upon them. A sphere of freely falling particles will be focused into an ellipsoid with the same volume, because the particles in the front of the sphere will fall faster than those in the rear, while at the same time the lateral cross-section of the sphere will shrink due to the tidal effect. This effect is responsible for the
gravitational breakup of very nearby massive stars.

Since gravitating matter is distributed inhomogeneously (except on the largest scales) causing inhomogeneous gravitational fields, Einstein realized that
the space we live in had to be curved, and the curvature had to be related to the distribution of matter. He then proceeded to search for a law of gravitation that was a generally covariant relation between mass density, as implied by the SEP, and curvature.
The simplest form for such a relation is \textit{Einstein's Equation}
\begin{equation}
G_{\mu\nu}=\frac{8\pi G}{c^4}T_{\mu\nu}.
\end{equation}
The \textit{Einstein tensor} $G_{\mu\nu}$ contains only terms which are either quadratic in the first spacetime derivatives of the metric tensor $g_{\mu\nu}$, or linear in the second derivatives. (Higher-order derivatives are difficult to include without making the theory unstable.) The \textit{stress--energy tensor} $T_{\mu\nu}$ contains the various components of energy densities, pressures and shears of matter and radiation.

The Einstein tensor vanishes for flat spacetime and in the absence of matter and pressure, as it should. Thus the problems
encountered by Newtonian mechanics have been resolved. The
recession velocities of distant galaxies do not exceed the speed of light, and effects of gravitational potentials are
not felt instantly. The discontinuity of homogeneity and isotropy at the boundary
of the Newtonian universe also disappeared because four-space is unbounded, and because spacetime in general relativity (GR) is generated by matter and pressure. Thus spacetime itself ceases to exist where matter does not exist, so there cannot be any boundary between an `inside' homogeneous universe and an `outside' spacetime void.

Einstein published his General Theory of Relativity in 1917, but the only solution he found to the highly nonlinear
differential equations (13) was static. This was in good agreement with the then known
Universe which comprised only the `fixed' stars in our Galaxy, and some nebulae of ill-known
distance and of controversial nature.

\section{Tests of General Relativity}
The classical testing ground of theories of gravitation, Einstein's among them, is celestial mechanics within the Solar
System.
The earliest phenomenon requiring general relativity for its explanation was noted in 1859, 20~years before Einstein's
birth. The French astronomer \textit{Urban Le Verrier} (1811--1877) found that
the planet Mercury's elongated elliptical orbit precessed slowly around the Sun. As the innermost planet it feels the solar
gravitation very strongly, but the orbit is also perturbed by the other planets. The total effect is that the \textit{perihelion} of the orbit advances 574$''$ (seconds of arc) per century. This is calculable using Newtonian mechanics and Newtonian gravity, but the result is 43$''$ too little.

With the advent of general relativity the calculations could be remade. This time the discrepant 43$''$ were
successfully explained by the new theory, which thereby gained credibility. This counts as the first one of three `classical' tests of GR.

The second classical test was the predicted deflection of a ray of light passing near the Sun. We shall come back to
that test in Section~7 on gravitational lensing. The third classical test was the gravitational shift of atomic spectra: the frequency of emitted radiation makes atoms into clocks which run slower in a strong
gravitational field. This was first observed in a cloud of plasma ejected by the Sun to an elevation of about
72\,000~km above the photosphere and  an effect only slightly larger than that predicted by GR was found.
Similar measurements have been made of radiation from the surface of more compact stars such as Sirius' companion, the white dwarf Sirius B.

A fourth test is based on the prediction that an electromagnetic wave suffers a time delay when traversing an increased gravitational
potential. It was carried out in 1971 with the radio telescopes at the Haystack and Arecibo observatories by emitting radar
signals towards Mercury, Mars and, notably, Venus, through the gravitational potential of the Sun. The round-trip time
delay of the reflected signal was compared with theoretical calculations. Further refinement was achieved later by
posing the Viking Lander on the Martian surface and having it participate in the experiment by receiving and
retransmitting a radio signal from Earth. This experiment found the ratio of the delay observed to the delay
predicted by GR to be $1.000\pm 0.002$.

The most important tests of GR have been carried out on the radio observations of pulsars that are members of binary pairs, notably the PSR~$1913+16$, a pair of rapidly rotating, strongly magnetized neutron stars discovered in 1974 by \textit{R.~A. Hulse\index{Hulse, R.~A.}} and \textit{J.~H. Taylor\index{Taylor, J.~H.}}, awarded the Nobel prize in 1993. If the magnetic dipole axis does not coincide with the axis of rotation (just as is the case
with Earth), the star would radiate copious amounts of energy along the magnetic dipole axis. These beams at radio
frequencies precess around the axis of rotation like the searchlights of a beacon. As the beam sweeps past our line of
sight, it is observable as a pulse with the period of the rotation of the star. Pulsars are the most stable clocks known in the Universe, the variation is about
$10^{-14}$ on timescales of 6--12~months. The reason for this stability is the intense self-gravity of a neutron star, that makes it almost undeformable until the very last few orbits when the binary pair coalesces into one star.

This system does not behave exactly as expected in Newtonian mechanics, hence the deviations provide several independent confirmations of GR. The largest relativistic effect is the apsidal motion of the orbit which is analogous to the advance of the perihelion of Mercury. A second effect is the counterpart of the relativistic
clock correction for an Earth clock. The travel time of light signals from the pulsar through the gravitational potential of its companion provides a further effect.

The slowdown of the binary pulsar
is indirect evidence that this system loses its energy by radiating gravitational waves.
Such waves travel through spacetime with the speed of light, traversing matter unhindered and unaltered, and producing ripples of curvature, oscillatory stretching and
squeezing of the web of spacetime analogously to the tidal effect of the Moon on Earth. Any matter they pass through will feel this
effect. Thus a detector for gravitational waves is similar to a detector for the Moon's tidal effect, but the waves act on an exceedingly weak scale.

In GR, the inertial and centrifugal forces felt on Earth are due to our accelerations and rotations with respect to the local inertial frames which, in turn, are determined, influenced and \textit{dragged} by the distribution and flow of mass densities in the Universe. A spinning mass will `drag' inertial frames and gyroscopes along with it. This also influences the flow of time around a spinning body, so that synchronization of clocks around a closed path near it is not possible. This effect is predicted by GR to be quite small.

Note that the expansion of the Universe and Hubble's linear law (1) are not tests of GR. Objects observed at wavelengths ranging from radio to gamma rays are close to isotropically distributed over the sky.
Either we are close to a center of spherical symmetry---an anthropocentric view---or the Universe is close to
homogeneous. In the latter case, and if the distribution of objects is expanding so as to preserve homogeneity and
isotropy, the expansion velocities satisfy Hubble's law.

\section{Gravitational lensing}
A consequence of SEP is that a photon in a
gravitational field moves as if it possessed mass, and light rays therefore bend around gravitating masses. Thus celestial bodies can serve
as \textit{gravitational lenses}. Since photons are neither emitted nor absorbed in the process of gravitational light deflection, the surface brightness
of lensed sources remains unchanged. Changing the size of the cross-section of a light bundle only changes the flux observed from a source and magnifies it at fixed surface-brightness level. For a large fraction of
distant quasars the magnification is estimated to be a factor of~10 or more which makes objects of faint intrinsic
magnitudes visible. If the mass of the lensing object is very small, one will merely observe a magnification of the brightness of the
lensed object. This is called \textit{microlensing}, and it has been used to search for nearby nonluminous objects in the halo of our Galaxy. Microlensing of distant quasars by compact lensing objects (stars, planets) has also been observed and used for estimating the mass distribution of the lens--quasar systems.

\textit{Weak Lensing} refers to deflection through a small angle. when the light ray can be treated as a straight line, and the deflection as if it occurred discontinuously
at the point of closest approach (the thin-lens approximation in Optics). One then only
invokes SEP which accounts for the distortion of clock rates. In \textit{Strong
Lensing} the photons move along geodesics in a strong gravitational potential which distorts space as well as time, causing larger deflection angles and requiring the full theory of GR. The images in the observer plane can then
become quite complicated because there may be more than one null geodesic connecting source and observer; it may not even be possible to find a unique mapping onto the source plane. Strong lensing is a tool for testing the
distribution of mass in the lens rather than purely a tool for testing GR. An important application of strong lensing is to measure cosmological distances and the global expansion rate $H_0$ by the Refsdal method (Section 5).

At cosmological distances one may observe lensing by composed objects such as galaxy groups which are ensembles of `point-like', individual galaxies. Lensing effects are very model-dependent, so to learn the true magnification effect one
needs very detailed information on the structure of the lens.

The observation of the deflection of light from stars at lines of vision near the Sun is an example of weak lensing. The Sun must then be fully eclipsed by the Moon to shut out its intense direct light, and the stars must be very bright to be visible
through the solar corona. Soon after the publication of GR in 1917 it was realized that such a fortuitous occasion to test the theory would occur. The Royal Astronomical Society then sent out two
expeditions to the path of the eclipse, one to Sobral in North Brazil and the other one to the Isle of Principe in the Gulf of Guinea. Both expeditions successfully observed several stars at various angles of deflection around the eclipsed Sun which confirmed the prediction of GR with reasonable confidence and excluded the Newtonian value. Similar measurements have been repeated many times since then during later solar eclipses, with superior results confirming the prediction of GR.

The large-scale distribution of matter in the Universe is inhomogeneous in every direction, so one can expect that
everything we observe is displaced and distorted by weak lensing. Since the tidal gravitational field, and thus the
deflection angles, depend neither on the nature of the matter nor on its physical state, light deflection probes the
total projected mass distribution. Lensing in infrared light offers the additional advantage of
being able to sense distant background galaxies, since their number density is higher than in the optical range.

Background galaxies would be ideal tracers of distortions if they were intrinsically circular, because lensing transforms circular sources into ellipses. Any measured ellipticity would then directly reflect the action of the gravitational tidal field of the interposed lensing matter, and the statistical properties of the distortions would reflect the properties of the matter distribution. But many galaxies are actually intrinsically elliptical, and the ellipses are randomly oriented. This introduces noise into the inference of the tidal field from observed ellipticities. A useful feature in the sky is a fine-grained pattern of faint and distant blue galaxies appearing as a `wall paper'. This makes statistical weak-lensing studies possible, because it allows the detection of the coherent distortions imprinted by gravitational lensing on the images of the galaxy population. Thus weak lensing is becoming an important technique to map non-luminous matter.

\section{Black holes}
Consider a body of mass $m$ and radial velocity $v$ attempting
to escape from the gravitational field of a spherical star of mass $M$. To succeed, the kinetic energy must overcome the gravitational
potential which varies with elevation. In Newtonian mechanics the condition for this is
\begin{equation}
\frac12mv^2\geq GMm/r.
\end{equation}
The larger the ratio $M/r$ of the star, the higher is the velocity required to escape. Ultimately, in the ultra-relativistic case when $v=c$, a non-relativistic treatment is no longer justified. Nevertheless, it just so happens that the equality in (13) fixes the radius of the star correctly to be the \textit{Schwarzschild radius}
\begin{equation}
r_{\rm c}\equiv\frac{2GM}{c^2},
\end{equation}
Because nothing can escape the interior of this event horizon $r_{\rm c}$, not
even light, \textit{John~A. Wheeler} coined the term \emph{black hole} for it in 1967.

The black hole is described by the \textit{Schwarzschild metric}
\begin{equation}
{\rm d}\tau^2=\bigg(1-\frac{r_{\rm c}}{r}\bigg)\,{\rm d} t^2-\bigg(1-\frac{r_{\rm c}}{r}\bigg)^{-1}\frac{{\rm d} r^2}{c^2}+\frac{r^2}{c^2}({\rm d}\theta^2+
\sin^2\theta\,{\rm d}\phi^2)\, ,
\end{equation}
which has very fascinating consequences. Consider a spacecraft approaching a black hole with apparent velocity $v={\rm d} r/{\rm d} t$ in the fixed frame of an outside observer. Light signals from the spacecraft travel on the light cone, ${\rm d}\tau=0$, so that
${\rm d} r/{\rm d}t=c(1-r_{\rm c}/r)$.
Thus the spacecraft appears to slow down with decreasing~$r$, finally coming to a full stop as it reaches $r_{\rm c}$.

The time intervals ${\rm d}t$ between successive crests in the wave of the emitted light become longer, reaching infinite wavelength at the mathematical
singularity of ${\rm d}t$ in the second term of Equation~(16). Thus the frequency $\nu$ of the emitted photons redshift to zero, and the energy $E=h\nu$ of
the signal vanishes. One cannot receive signals from beyond the event horizon because photons cannot have negative energy.

To the pilot in the spacecraft the singularity at $r_{\rm c}$ does not exist, his comoving clock shows finite time when he reaches the event horizon and crosses it.
At the center of the black hole the metric has a physical singularity where GR breaks down. Some people have speculated that matter or radiation falling in might `tunnel' through a `wormhole' out into another universe.

\textit{J. Bekenstein} noted in 1973 that there are certain similarities between the
surface area $A$ of the event horizon of a black hole and entropy $S$. When a star has collapsed to the size of its Schwarzschild radius, its event horizon will never change (to an outside observer) although the collapse continues. Thus entropy could be defined as $S\propto A$. The area can increase only if the black hole devors more mass from the outside, but $A$ can never decrease because no mass will leave the horizon.

\textit{Stephen Hawking} has shown that black holes can
nevertheless radiate if one takes quantum mechanics into account. It is a property of the vacuum that particle--antiparticle pairs such as ${\rm e^-e^+}$ are continuously created out of nothing, to disappear in the next moment by annihilation which is the inverse
process. Since energy cannot be created or destroyed, one of the particles must have positive energy, and the other an equal amount of negative energy. They form a virtual pair, neither one is real in the sense that it could escape to infinity or be observed by us.

However, in a strong electromagnetic field the ${\rm e^-}$ and the ${\rm e^+}$ may become separated by a Compton wavelength $\lambda\approx r_{\rm c}$. Then there is a small but finite probability for one of them to `tunnel' through the barrier of the quantum vacuum and escape the black
hole horizon as a real particle with positive energy, leaving the negative-energy particle inside the horizon of the hole. Since energy must be conserved, the hole loses mass in this process, a phenomenon called \textit{Hawking radiation}.

Black holes manifest their presence within many galaxies through their gravitational action on surrounding stars, and through powerful jets of radiation which cannot be fueled by any other phenomena. The center of the Milky Way hosts a super-heavy black hole with a mass of several million solar masses. Black holes may have been created in the Big Bang, and they are created naturally in the ageing of heavy stars. Black holes grow by accreting surrounding matter, and by mergers with neighboring black holes. Gravitational merger waves from super-heavy black holes may be so strong that both their direction and their amplitude can be monitored by detectors on satellites, permitting tests of GR and determination of fundamental cosmological parameters of the Universe.

The Schwarzschild metric has been quantitatively verified in the weak-field limit on small scales such as the Solar system, in the binary radio pulsars, and on Galaxy scales.

      \section{Friedmann--Lema\^{\i}tre cosmology}

Immediately after General Relativity became known, \textit{Willem de~Sitter}
(1872--1934) found and published an exponentially expanding solution to Einstein's
equations (13) for the special case of empty spacetime. In 1922 \textit{Alexandr Friedmann} (1888--1925) found a range of solutions, intermediate to Einstein's static solution and de~Sitter's solution.  Friedmann's solutions did not gain general recognition until after his death when they were confirmed by an independent derivation (in 1927) by
Georges Lema\^{\i}tre (1894--1966). The latter is considered to be the father of the Big Bang model rather than Friedmann. Only in 1934 did Robertson and Walker construct the RW metric (8) to match the general geometrical structure of the Einstein's tensor $G_{\mu\nu}$.

Today the standard model of cosmology is based on the Friedmann--Lema\^{\i}tre equations (FL) and the RW metric. For a comoving observer with velocity four-vector $v=(1,0,0,0)$ in a homogeneous and isotropic Universe only two of the 16 differential equations in the tensor equation~(13) are needed:
\begin{equation}
\left(\frac{\dot a}{a}\right)^2\equiv H^2= -\frac{kc^2}{a^2}+\frac{8\pi G}{3}\rho, ~~~~~~~~~
\frac{2\ddot{a}}{a}+H^2=-\frac{kc^2}{a^2}-\frac{8\pi Gp}{c^2}.
\end{equation}

The expansion or contraction of the Universe is inherent to these equations. The first equation shows that the rate of expansion, $\dot{a}$, increases with the density of matter, $\rho$, and the second equation shows that $\rho$ and the pressure, $p$, decelerate the expansion. Eliminating $\ddot{a}$ from the second equation by using the first, one obtains the covariant \emph{conservation equation} of the energy-momentum tensor in an homogeneous and isotropic spacetime,
\begin{equation}
\dot{\rho}+3H(\rho+pc^{-2})=0.
\end{equation}
Assuming that $\rho$ and $p$ are linearly related by an \emph{equation of state}, $w\equiv p/\rho c^2$, one conveniently replaces $(\rho +pc^{-2})$ by $\rho\,(1+w)$.

Equation~(18) is the tool to find the $a$-dependence of an energy density. In the present matter-dominated universe filled (to a good approximation) with non-relativistic
cold, pressureless, non-radiating dust, one has $w=0$, and the density evolves as $\rho_{\rm m}(a)\propto a^{-3}=(1+z)^3$.

In an earlier radiation-dominated universe filled with an
ultra-relativistic hot gas composed of elastically scattering particles, statistical
mechanics tells us that the equation of state was $p_{\rm r}=\rho_{\rm r}/3$. This corresponds to $w=1/3$, so that the radiation density evolves as
$\rho_{\rm r}(a)\propto a^{-4}=(1+z)^4$

The first equation~(17) can easily be integrated to give the scale-dependence of time, $t(a)$. However, to solve by algebraic methods the time-dependence of the scale, $a(t)$, is only possible in flat space when $k=0$. During matter-domination the scale evolves as $a(t)\propto t^{2/3}$, during radiation-domination as $a(t)\propto t^{1/2}$.
The two scales must then have been equal at some specific time in the past, the time of \emph{matter-radiation equality}, $t_{\rm eq}\approx 60\,000$ years after Big Bang.

Note that as one approaches $t\approx 0$, also the scale goes to zero, and the radiation energy density approaches infinity. This singularity is termed the \emph{Big Bang}. Intuitively it is a meaningless and unphysical result, given the knowledge that quantum mechanics does not allow an exactly zero scale nor time. But it is an understandable result, because GR does not contain quantum mechanics, so it must break down as a description when the size of the Universe approaches atomic scales. A better future theory must combine GR and quantum mechanics.

In the static universe of Einstein $a(t)$ is constant and the age of the Universe is infinite. In order that $\rho_0$ today be positive, $k$ must then be~$+1$. This leads to the surprising result that the Universe is contracting and $p_0$ negative. Einstein did not believe this to be true, so he corrected Equation~(13) by introducing a constant Lorentz-invariant term $\Lambda g_{\mu\nu}$. The \textit{cosmological constant} $\Lambda$ corresponds to a
tiny correction to the geometry of the Universe which adds enough repulsion to make the Universe static. Unfortunately Einstein was wrong, Hubble showed that it was expanding, and Einstein admitted $\Lambda$ had been a blunder. However, as we shall see, $\Lambda$ has been resurrected since then, but not in order to make the Universe static.

The cosmological constant can be considered as a mere classical modification of Newton's gravitational constant $G$, but alternatively as a contribution from a vacuum energy term $-\Lambda g_{\mu\nu}T_{\mu\nu}/8\pi G$ on the right-hand side of Einstein's equations. The two interpretations affect the geometry of spacetime in exactly the same way, and are thus observationally indistinguishable. Vacuum energy can be physically interpreted as an
ideal fluid with energy density $\rho_{\Lambda}=\Lambda/8\pi G$ and negative pressure $p_{\Lambda}=-\rho_{\Lambda}c^2$. The gravitational effect of a positive $\Lambda$ is a universal cosmic repulsion counteracting the attractive gravitation of matter. The vacuum energy state and the cosmological constant both have an equation of state parameter $w=-1$.

It is convenient to normalize all densities to the \emph{critical
density} $\rho_c=3H_0^2/8\pi G$, and to replace them by the dimensionless density parameters
$\Omega_m(a),~\Omega_r(a),~\Omega_{\Lambda},~\Omega_0\equiv\rho_0/\rho_c$, and possibly others. The sum of density parameters then add up to the present value $\Omega_0$. One can
also make the replacement $\Omega_k=1-\Omega_0$, where
$\Omega_k=kc^2H_0^{-2}$ is the vacuum energy density. From the first
FL equation (17) one sees that the Hubble `constant' $H$
actually is a function of the scale $a=1/(1+z)$. Making use of the $a$-dependence of the different density components we can write
\begin{equation}
H(a)=H_0~a\,(\Omega_{\rm k}a^{-2}+\Omega^0_{\rm m}a^{-3}+\Omega^0_{\rm
r}a^{-4}+\Omega_{\Lambda})^{1/2}.
\end{equation}
The function $H(a)$ can be used to determine the age of the Universe by integrating $H(a)^{-1}$ from $a=0$ to $a=1$, or the \emph{lookback time} by integrating from the present time $t_0$ to the time $t(a)$ at scale $a$.

The case $k=\Lambda=0$ is called the \emph{Einstein--de Sitter universe}. An
already mentioned special case is the empty \emph{de Sitter universe} which expands exponentially with constant acceleration. Although the present Universe is not empty it does accelerate, making this model interesting: in a distant future an ever expanding universe indeed approaches emptiness, since the matter density is diluted as $\rho_{\rm m}(a)\propto a^{-3}$.

\section{Thermal History of the Universe}

The thermal history of the Universe begins soon after the Big Bang when the
Universe was in a state of extreme heat and pressure, occupying an exceedingly small volume in the era of \textit{radiation
domination}. From the time when the mean temperature was $T=10^{16}$~K (in energy units $E=kT\approx1$~TeV where $k$ is the Boltzmann constant), and the scale was $a\approx10^{-16}$, our model of the Universe is fairly reliable, because this is the limit of present-day laboratory experimentation and the standard model of particles. At earlier times new physics may appear, such as \textit{supersymmetry} (SUSY).

As the Universe cooled, the primeval plasma condensed to quarks, gluons, vector bosons (W, Z), leptons (electrons, muons, neutrinos), baryons (\emph{e.g.} nucleons), and various unstable mesons. Their number of spin states -- even for fermions, odd for bosons, 2 for the photon -- determines their degrees of freedom and how they behave in a statistical ensemble.

All particles were relativistic, they were incessantly colliding and
exchanging energy and momentum with each other and with the radiation photons. A few collisions were sufficient to establish a \textit{thermal equilibrium} in which the available energy was distributed evenly among all particles in a stable energy spectrum, the \textit{blackbody spectrum}. This spectrum is completely characterized by only one parameter, its temperature $T$.

Which particles participate in thermal equilibrium at a given energy depends on two timescales: the reaction rate of the particle, and the expansion rate of the Universe. If the reaction rate is slow compared to the expansion rate, the distance between particles grows so fast that they cannot find each other, and those particles drop out of thermal equilibrium. For much of the thermal history of the Universe, the reaction rates of photons and other particles have been much greater than the Hubble expansion rate, so thermal equilibrium should have been maintained in any local comoving volume element. There is then no net inflow or outflow of energy, which defines the expansion as \emph{adiabatic}. In a thermodynamic context, conservation of energy of adiabatically expanding matter is embodied by the \textit{first law of thermodynamics}.

The \emph{second law of thermodynamics} states that any isolated system left by itself can only change towards
greater disorder or \emph{entropy}. The counterexample which living organisms seem to furnish since they build up ordered systems is not valid, because no living organism exists in isolation: it consumes ordered nutrients and produces disordered waste.

Independently of the first law, the second law implies that energy is distributed equally between all \textit{} present. A change in degrees of freedom is accompanied by a change in random motion, or equivalently in temperature. The more degrees of freedom there are present, the more randomness or disorder the system possesses. Molecules in a gas possessing kinetic energy only (heat) attain maximal entropy when thermal equilibrium is reached. For a system of gravitating bodies entropy increases by clumping; maximal entropy corresponding to a black hole.

When the temperature had fallen to $10^{11}$K (or $kT\approx300$~MeV) all electrons and photons had an energy below the threshold for ${\rm p}\bar{\rm p}$ production. The number of baryons then no longer increased as a result of thermal collisions; their number densities decreased exponentially because they had become non-relativistic. They rather annihilated into lepton pairs, pion pairs or photons.  If there had been exactly the same number of nucleons and anti-nucleons, we would not expect many nucleons to remain to form matter. But, since we live in a matter-dominated Universe, there must have been some excess of nucleons early on. The cause of this primordial excess is not yet understood despite much effort.

Below 70 MeV the temperature in the Universe cooled below the threshold for pion
production, and below 50 MeV for muon-pair production. Pions and muons decay in a time much shorter than the age of the Universe, then about 1~ms. After that we are left with relativistic electrons and neutrinos and non-relativistic nucleons which still participated in thermal equilibrium, although they were too few to play any role in the thermal history. At 2.3 MeV the lightest neutrinos \emph{decoupled} from all interactions, and began a free expansion.

As long as there were free electrons, the primordial photons were thermalized by elastic scattering against them, and this coupled the electrons and photons. Electrons only decoupled when they formed neutral atoms with protons during the \textit{recombination era} and ceased to scatter photons. Photons with energies below the electron mass could no longer produce ${\rm e}^+ -{\rm e}^-$ pairs, but the energy exchange between photons and electrons still continued by elastic (Compton or Thomson)  scattering. The annihilation reaction ${\rm e}^+ +{\rm e^-}\rightarrow\gamma~\gamma$  was now of mounting importance, creating new photons with energy 0.51 MeV. This was higher than the ambient photon temperature at that time, so the photon population got reheated by a factor 1.40.

From the time $t_{eq}$ (or $kT\approx1$~eV) when the temperatures of matter and radiation were equal, cold matter and hot radiation decoupled, matter cooling faster than radiation.  Thus they were no longer in thermal equilibrium and will never be so on a cosmic timescale in an expanding Universe.

After recombination, density perturbations in matter were no longer damped by interaction with radiation, so they could grow into structures through gravitational instability. Decoupling thus initiates the period of structure formation, that has led to our present Universe being populated with stars, galaxies, and galaxy clusters.

A second consequence of decoupling was that, with photons no longer scattering against a sea of electrons, they could stream freely through the Universe; upon recombination, the Universe became transparent to light. Prior to recombination, the Universe was opaque to electromagnetic radiation (although not to neutrinos) and it would have been impossible to do astronomy if this had persisted until today. The freely streaming photons continued as microwaves and infrared light from their last point of contact with matter on a spherical shell called the \textit{last scattering surface} (LSS). The era of recombination provides a crucial observational limit beyond which we cannot hope to see using electromagnetic radiation.

Already at a few MeV, nuclear \textit{fusion reactions} started to build light elements. In reactions when a neutron and a proton fuse into a bound state, some of the nucleonic matter is converted into pure energy according Equation~(6).  As the Universe cooled the neutron-producing reactions stopped, some neutrons got converted into protons in various reactions, until the neutron/proton ratio had been reduced from 1 to 1/7. At this time the energy was 0.1~MeV and the time elapsed since the Big Bang a little over two minutes.

The remaining neutrons had no time to decay before they fused into deuterons and subsequently via a chain of reactions into ${}^4$He, where they stayed until today because bound neutrons do not decay. The same number of protons went into ${}^4$He, and the excess became the nuclei of hydrogen atoms. Thus the end result of this \textit{nucleosynthesis} taking place 100--700~s after the Big Bang is a Universe composed almost entirely of hydrogen and helium. The best information today on the primordial baryonic density comes from the observed deuterium abundance.

Big Bang cosmology makes some very important testable predictions:
\begin{itemize}
\item
The Universe today should still be filled with freely streaming relic photon (and neutrino) radiation with a blackbody spectrum of temperature related to the age of the Universe and a polarization correlated to the temperature. \item
This relic radiation should be essentially isotropic since it originated in the now spherical shell of the LSS. In particular, it should be uncorrelated with the radiation from foreground sources of later date, such as our Galaxy.
\end{itemize}

\section{Cosmic inflation}

A particle horizon exists if the age of the Universe is finite. The particle horizon and the spatial width of the past light cone have grown in proportion to the longer time perspective. Thus the spatial extent of the Universe is larger than what our past light cone encloses today; with time we will become causally connected with new regions as they move in across our horizon. This renders the question of the full size of the whole Universe meaningless -- the only meaningful size being the diameter of its horizon at a given time.

We have argued that thermal equilibrium could be established throughout the Universe during the radiation era because photons could traverse the whole Universe and interactions could take place in a time much shorter than a Hubble time, $H^{-1}$. However, there is a snag to this argument: the conditions at any spacetime point can only be influenced by events within its past light cone, the size of which at the time of last scattering
($t_{\rm LSS}$) was far too small to allow the currently observable Universe to come into thermal equilibrium. Comoving bodies at the Hubble radius recede with the velocity $c$, but the particle horizon, $\sigma_{\rm ph}(k,\chi_{\rm ph})$ (see Equation~(9)), recedes even faster. The net effect is that today's particle horizon covers regions which were causally disconnected in the radiation era.

The event horizon at the time of last scattering, $\sigma_{\rm eh}(t_{\rm LSS})$, represents the observable extent of the Universe today (cf. Figure~1).
On the other hand $\sigma_{\rm ph}(t_{\rm LSS})$ represents the extent of the LSS that could have been in causal contact from $t=0$ to $t_{\rm LSS}$. If $\sigma_{\rm eh}(t_{\rm LSS})>\sigma_{\rm ph}(t_{\rm LSS})$, the visible Universe could not have been causally connected at time $t_{\rm LSS}$. Today $\sigma_{\rm ph}(t_{\rm LSS})$ is seen
as an arc on the periphery of our particle horizon, subtending an angle corresponding to the fraction $\sigma_{\rm ph}/\sigma_{\rm eh}$ or $1.12^{\circ}$. This is referred to as the \emph{horizon problem}.

Another problem is flatness. From the $a$-dependence of $\Omega_r$
in the era of radiation domination (when $\Omega_m$ and
$\Omega_{\Lambda}$ could be neglected) one has $(\Omega_0-1)\propto
a^2$. Thus when $a<10^{-16}$ the Universe must have been flat to
within $>32$~decimal places, a totally incredible situation. If this
were not so the Universe would either have reached its maximum size
within $<10^{25}$s and thereafter collapsed into a singularity, or
it would have dispersed into a vanishingly small energy density.
This is referred to as the \emph{flatness problem}.

Even more serious problems emerge as we approach $kT=11$~TeV, beyond which new physics is expected to occur (the Grand Unification Theory, \emph{GUT}). The temperature of the cosmic background radiation was then $\simeq4.4\times10^{27}$ times higher than today, and the linear scale $a(t)$ was smaller by the same factor. Note, however, that linear size and horizon are two different things. If we take the present Universe to be of size $2000h^{-1}$~Mpc, its linear size was only 2 cm at GUT time. To arrive at the now homogeneous Universe, the homogeneity at GUT time must have extended out to a distance $5\times10^{26}$ times greater than the
distance of causal contact! Why did the GUT phase transition happen simultaneously in a vast number of causally disconnected regions? Concerning even earlier times, one may ask the same question about the Big Bang.

The solution is to provide a mechanism which blows up the Universe so
rapidly, and to such an enormous scale, that the causal connection between its different parts is lost, yet they are similar due to their common origin. This should solve the horizon problem and flatten the local fluctuations to near homogeneity. There now exists a vast number of \emph{inflation} models which assume, that the $r_{\rm Pl}$-sized universe at Planck time $t_{\rm Pl}$ was pervaded by a homogeneous scalar classical field $\varphi$, the \textit{inflaton} field, and that all points in that primordial universe were causally connected. This idea is already embedded in the second FL equation~(17): if the pressure $p$ is negative, as in the presence of a cosmological constant, the Universe can accelerate. Inflationary models can be constructed in other ways, too, by increasing the dimensionality of spacetime and replacing pointlike particles by \emph{strings} in higher dimensions, by modifying Einstein's geometry $G_{\mu\nu}$, etc.

Clearly the cosmic inflation cannot go on forever if we want to arrive at our present slowly expanding FL universe. Thus there must be a mechanism to halt the exponential expansion, a \textit{graceful exit} after about 65 e-foldings, enough to solve the flatness problem. At that time the inflation has lasted $65\times10^{-34}$~s,
the initial particle horizon has been blown up by a factor of $10^{29}$, and thereby also the horizon problem is solved. At the end of inflation the Universe is an expanding hot plasma bubble of radiation and particles-to-be in thermal equilibrium. The latent heat stored as vacuum energy is liberated in the form of radiation, kinetic energy, and entropy of ultrarelativistic massive scalar particles with positive pressure. The energy density term in Friedmann's equations has become dominant, and the Universe henceforth follows an FL-type evolution.

Even if the Universe at Planck time was empty, Quantum Field Theory
tells us that empty space is filled with zero-point quantum
fluctuations, here fluctuations of the inflaton field. When they had
reached the length scale $H^{-1}$, they froze to an average
nonvanishing amplitude, and the vacuum no longer appeared empty and
devoid of properties. Fluctuations on longer scales were expelled
outside the present causal horizon so they no longer communicated,
crests and troughs in each oscillation mode remained frozen. But
after the end of inflation the expansion slowly returned these
frozen fluctuations inside the horizon. With time they became the
seeds of the perturbations we now observe in the Cosmic
Microwave Background Radiation (CMBR) and in the density distribution of
matter.

Inflationary cosmologies make several falsifiable predictions which
appear to hold rather well:
\begin{itemize}
\item \textit{Adiabaticity:} The fluctuations in the local number density of each species of matter should be the same and coupled to those of radiation;
\item \textit{Near scale invariance:} When the
matter fluctuations cross the Hubble radius their amplitudes should be nearly equal on all scales. This produces a power spectrum of density perturbations of the \emph{Harrison--Zel'dovich} form $P(k)\propto k^n$, where the \emph{scalar spectral index} is $n\approx1$;
\item \textit{Gaussianity:} Fluctuations in the CMBR and in the late-time evolution of large-scale structure should be Gaussian.
\item \textit{Curvature:} Flat space, $k=0,~~~~\Omega_0=1$.
\end{itemize}

\section{Cosmic Microwave Background Radiation}

The tight coupling between radiation and matter density before decoupling
caused the adiabatic perturbations to oscillate in phase.
Beginning from time $t_{\rm LSS}$, the receding horizon has been revealing these
frozen density perturbations, setting up a pattern of standing
acoustic waves in the baryon--photon fluid.  After
decoupling, this pattern is visible today as temperature anisotropies with a certain regularity across the sky.

The number of photons is now conserved, there are no creation nor
annihilation processes producing new photons, the Universe is
transparent to radio waves, and there is no known mechanism that
could recently have produced a blackbody spectrum in the microwave
range -- later radiation from astronomical bodies is much hotter.
The relic radiation which had a temperature $T$ at LSS time $t$
should still be blackbody at time $t'$ when the temperature has
scaled to $T'=T\frac{a(t)}{a(t')}$ in the microwave range. Thus the
Cosmic Microwave Background Radiation (CMBR) which has existed since the era of radiation domination could be
predicted, and was indeed predicted in 1948 at almost the right
temperature by \textit{George Gamow} (1904--1968), \textit{Ralph
Alpher} (1921--2007) and \textit{Robert Herman} (1914--1997).

In 1964 \textit{Arno Penzias} (1933-- ) and \textit{Robert Wilson}
(1936-- ) were testing a sensitive antenna intended for satellite
communication  They did not know about the predicted CMBR, they only
wanted to calibrate their antenna in an environment free of all
radiation, so they chose a wavelength of $\lambda=0.0735$ m in the
relatively quiet window between the known emission from the Galaxy
at longer wavelengths and the shorter wavelengths from the Earth's
atmosphere. They also directed the antenna high above the galactic
plane, where scattered radiation from the Galaxy would be minimal.

To their consternation and annoyance they found a constant low level
of background noise in every direction, that did not seem to
originate from distant galaxies since they did not see an intensity
peak in the direction of the nearby M31 galaxy in Andromeda. It
could also not have originated in Earth's atmosphere, because such
an effect would have varied with the altitude above the horizon as a
function of the thickness of the atmosphere.

This was the serendipitous discovery of the 15-Gyr-old echo of the
Big Bang, the most important discovery in cosmology since Hubble's
law. It is hard to understand how the
CMBR with a blackbody spectrum could have arisen without the cosmic material having
once been highly compressed, exceedingly hot, and in thermal equilibrium.

In 1993 detectors on the \emph{COBE} satellite verified the blackbody nature of the CMBR energy spectrum and saw the expected temperature fluctuations. Since then, these observations have been repeated by many other experiments with ever increasing precision. The temperature of this radiation is $T_0=2.725$~K with a precision of the blackbody curve far exceeding laboratory experiments.  Knowing $T_0$
one can calculate the present values of the energy density and
entropy density of radiation, as well as the neutrino number density
and temperature, $T_{\nu}=1.946$~K.

The spatial distribution of the CMBR temperature exhibits a \textit{dipole anisotropy}
($\ell=1$). It is maximally blueshifted in one direction of the sky
and maximally redshifted in the opposite direction. In the standard
model the cosmic expansion is spherically symmetric, and inflation predicts
that the CMBR should be isotropic, since it originated in the LSS,
which has now receded to a radial distance of $z\simeq1100$. Thus it
is quite clear that the dipole anisotropy cannot be of cosmological
origin. Rather, it is well explained by the motion of our Local
Group of galaxies `against' the radiation in the direction of
maximal blueshift. Anisotropies on the largest angular scales
corresponding to quadrupoles ($\ell=2$) would be manifestations of truly
primordial gravitational structures.

The primordial photons are polarized by the anisotropic Thomson
scattering process, but as long as the photons continue to meet free
electrons their polarization is washed out, and no net polarization
is produced. At a photon's last scattering, however, the induced
polarization remains and the subsequently free-streaming photon
possesses a quadrupole moment.

Temperature and polarization fluctuations around a mean temperature
$T_0$ in a direction $\alpha$ on the sky are analyzed in terms of
the \textit{autocorrelation function} $C(\theta)$, which measures
the product of temperatures in two directions,
$\alpha+\frac12\theta$ and $\alpha-\frac12\theta$, averaged over all
directions $\alpha$. The $C(\theta)$ are expanded in spherical harmonics of the observed temperature; the expansion coefficients are called powers or multipole components. The resulting distribution of powers versus multipole $\ell$, or multipole moment $k=2\pi/\ell$, is the \textit{power spectrum} which exhibits conspicuous \emph{Doppler peaks}. The spectrum can then be compared to theory, and theoretical parameters determined. Many experiments have determined the power spectrum, so that a wealth of data exists.

The six parameters that can be best determined by fitting the angular scale of the Doppler peaks in the spectrum of
temperature and polarization power spectra are the total matter
density, $\Omega_m$, the baryonic matter density, $\Omega_b$, the
Hubble parameter, $h$, the scalar spectral index, $n$, the amplitude of the
fluctuations, and the Thomson scattering optical depth to
decoupling. In addition there is some information on the sum of
neutrino masses, the equation of state parameter $w_{\Lambda}$ for
$\Omega_{\Lambda}$, and a value for $\Omega_{\Lambda}$ can be
derived from $\Omega_{\Lambda}=1-\Omega_m$ if one assumes flat space
($\Omega_k=0$). Equation~(19) can now be used to determine the age of the universe: $t_U=13.7\pm0.2$~Gyr which is in good agreement with cosmochronological determinations.

The results show two surprises: Firstly, $\Omega_m\approx 0.24\ll1$
so that a large component $\Omega_{\Lambda}\approx0.76$ is missing,
of unknown nature, and termed \textit{dark energy}. The second
surprise is that ordinary baryonic matter is only a small fraction
of the total matter budget, $\Omega_b\approx0.044\ll\Omega_m$. This
missing matter is termed \textit{dark matter}, of unknown composition.
Of the 4.4\% of
baryons in the Universe 80\% is gas, 20\% dust, 1\% stars. The
baryon/photon number ratio is $\approx6.3\times 10^{-10}$.

Some of these parameters are quite degenerate, so that one needs
to combine CMBR data with other data, notably with the \emph{baryonic acoustic oscillation} (BAO) scale detected in galaxy clustering statistics and type Ia supernovae luminosity distances.

\section{Large Scale Structure}

A cornerstone of cosmology is the Copernican principle, that matter in the Universe is distributed homogeneously, if only on the largest scales of superclusters separated by voids. On smaller scales we observe inhomogeneities in the forms of galaxies, galaxy groups, and clusters. The common approach to this situation is to turn to non-relativistic hydrodynamics and treat matter in the Universe as an adiabatic, viscous, non-static fluid, in which random fluctuations around the mean density appear naturally, manifested by compressions in some regions and rarefactions elsewhere.  The origin of these density fluctuations was the tight coupling
established before decoupling between radiation and charged matter density, causing them to oscillate in phase. An ordinary fluid is dominated by the material pressure, but in the fluid of our Universe three effects are competing: gravitational attraction, density dilution due to the Hubble flow, and radiation pressure felt by charged particles only.

Inflationary fluctuations will cross the post-inflationary Hubble radius and come back into vision with a wavelength corresponding to the size of
the Hubble radius at that moment. At time $\approx t_{eq}$ the overdensities begin gravitational amplification and grow into larger inhomogeneities.  In overdense regions where the gravitational forces dominate, matter contracts locally and attracts surrounding matter, becoming increasingly unstable until it eventually collapses into a gravitationally bound object. In regions where the pressure forces dominate, the fluctuations move with constant amplitude as sound waves in the fluid, transporting energy from one region of space to another.

Inflationary models predict that the primordial mass density fluctuations should be adiabatic, Gaussian, and exhibit the same scale invariance as the CMBR fluctuations.
The density fluctuations can be specified by the amplitudes $\delta_k$ of the dimensionless \textit{mass autocorrelation function}
$\xi({\bf r})=\langle\delta({\bf r_1})~\delta({\bf r}+{\bf r_1})\rangle$
which measures the correlation between the density contrasts at two points ${\bf r}$ and ${\bf r_1}$. The autocorrelation function is the Fourier transform of the power spectrum. In this way, baryonic acoustic oscillations in the mass density power spectrum can be related to the coefficients in a spherical harmonic expansion of $\xi({\bf r})$, similarly as in the CMBR case, and they can now serve to determine cosmological parameters of theoretical models.

As the Universe approaches decoupling, the photon mean free path increases
and radiation can diffuse from overdense regions into underdense ones, thereby smoothing out any inhomogeneities in the plasma. Similarly, free streaming of weakly interacting relativistic particles such as neutrinos erases
perturbations up to the scale of the horizon. The situation changes dramatically at recombination, when all the free electrons suddenly disappear, captured into
atomic Bohr orbits, and the radiation pressure almost vanishes. This occurs at time 380\,000~yr after Big Bang. Now the density perturbations which have entered the Hubble radius can grow with full vigor into baryonic structures.

The timetable for galaxy and cluster formation is restricted by two important constraints. At the very earliest, the Universe has to be large enough to have space for the first formed structures. Also, the density contrast at formation must have exceeded the mean density at that time. Since then the contrast has increased with $a^3$. Thus, rich clusters, for instance, cannot have been formed much earlier than at $z=5$.

Galaxy formation can be simulated because the only force is
gravitational and the dominant energy is non-interacting dark
matter. Star formation cannot be simulated as easily, because the
electromagnetic forces are important in addition to gravitation, and collisionless
(dark) matter plays no role. Hierarchical structure formation scenarios (galaxies fueled and growing by mergers, coalescence of black holes) predict a
decreasing luminosity, that is problematic because it is not supported by data up to z=2.

\section{Dark Matter}

That the dominant fraction of matter is missing, unknown, invisible and therefore dark (DM), has now been demonstrated in 10 different ways and on different scales.

\textbf{1.} Baryonic matter feels attractive self-gravity and is pressure-supported, whereas dark matter only feels attractive self-gravity, but is pressureless. Thus the Doppler peaks in the CMBR power spectrum testify about baryonic and dark matter, whereas the troughs testify about rarefaction caused by the baryonic pressure. The position of the first peak determines $\Omega_m h^2$. Combining this with determinations of the Hubble constant $h$, one finds the total mass density parameter $\Omega_m\approx0.24$. The ratio of amplitudes of the second/first Doppler peaks determines the baryonic density parameter $\Omega_b$, and the difference is $\Omega_{\rm dm}\approx 0.2$.

\textbf{2.} Baryonic acoustic oscillations also testify about DM. However, the scale of BAO depends on $\Omega_m$ and the Hubble constant, $h$, so one needs information on $h$ to break the degeneracy. The result is then $\Omega_m\approx0.26$.

\textbf{3.} If the galaxies had arisen from primordial density fluctuations in a purely baryonic medium, the amplitude of the fluctuations should have been very large, since $\Omega_b$ is so small. But the amplitude of
fluctuations in the CMBR would then also be very large, because they maintain adiabaticity. This would have lead to intolerably large CMBR anisotropies today. Thus galaxy formation in purely baryonic matter is ruled out by this argument alone.

\textbf{4.} Spiral galaxies are stable gravitationally bound systems in which visible matter is composed of stars and interstellar gas.
Most of the observable matter is in a relatively thin disc, where stars and gas rotate around the galactic center on
nearly circular orbits. If the circular velocity at
radius $r$ is $v$ in a galaxy of mass~$M$, the condition for stability is that the centrifugal acceleration should
equal the gravitational pull, and the radial dependence of $v$ would then be expected to follow Kepler's law
\begin{equation}
v=\sqrt{\frac{GM}{r}}.
\end{equation}
The surprising result from measurements of galaxy-rotation curves is that the velocity does not follow this $1/\sqrt{r}$ law, but stays constant after attaining a maximum at about 5~kpc. The most obvious solution to this is that the galaxies
are embedded in extensive haloes of DM, of the order of $\Omega_{\rm halo}>0.03-0.10$. This would make the radial mass distribution $M(r)$ proportional to $r$, so that $v(r)=$~constant.

\textbf{5.} Let us define the mass/luminosity ratio of an astronomical object as $\Upsilon\equiv M/L$. Stellar populations exhibit values $\Upsilon=1-10$, the Solar neighborhood $\Upsilon\approx2.5-7$, rich clusters on the largest scales $\Upsilon\geq300$. Since $L_{\rm Universe}$ is known, one derives that the critical density to close the Universe would require $\Upsilon\approx990$.
However, there is not sufficiently much luminous matter to achieve closure, a large amount of DM is needed.

\textbf{6.} There are examples of groups formed by a small number of galaxies
which are enveloped in a large cloud of hot gas, visible by its X-ray emission. The amount of gas can be deduced from the intensity of this radiation. Adding this to the luminous matter seen, the total amount of baryonic matter, $M_b$, can be estimated.
The temperature of the gas depends on the strength of the gravitational field, from which the total amount of gravitating matter, $M_{\rm grav}$, in the system can be deduced. In many such small galaxy groups one finds $M_{\rm grav}/M_b\geq3$. Thus a dark halo must be present.

\textbf{7.} Over large distances (tens of Mpc) where the peculiar velocities of galaxies are unimportant relative to the Hubble flow, the galaxies in the Local Supercluster (LSC) feel its attraction, as is manifested by their infall toward its center with velocities in the
range 150--450~km\,s$^{-1}$. From this one can derive the local gravitational field and the mass excess concentrated in the LSC.

If galaxy formation is a local process, then on large scales galaxies must trace mass. By comparing the mass autocorrelation functions $\xi_{\rm gal}(r)$ and $\xi_{\rm mass}(r)$, averaged over the LSC volume, one concludes that a large amount of matter in the LSC is dark.

\textbf{8.} \textit{Fritz Zwicky} (1898--1974) noted in 1933 that the galaxies in the Coma cluster and other rich clusters move so fast that the clusters
require about 10--100 times more mass to keep the galaxies bound than could be accounted for by the luminous mass in the galaxies themselves. This was the earliest indication of DM in objects at cosmological distances.

The virial theorem for a statistically steady, spherical, self-gravitating cluster of objects, stars or galaxies, states that the total kinetic energy of $N$ objects with average random peculiar velocities $v$ equals $-\frac{1}{2}$ times the total gravitational potential energy. If $r$ is the average separation between any two objects of average mass $m$,
the potential energy of each of the possible $N(N-1)/2$ pairings is $-Gm^2/r$. The virial theorem then states that
\begin{equation}
N\frac{mv^2}{2}=\frac{1}{2}\frac{N(N-1)}{2}\frac{Gm^2}{r}.
\end{equation}
One can apply the virial theorem to estimate the total dynamic mass of a rich galaxy cluster from measurements of the velocities of the member galaxies and the cluster radius from the volume they occupy. When such analyses have been carried out, taking into account that rich clusters have about as much mass in hot gas as in stars, one finds that
gravitating matter accounts for $\Omega_{\rm grav}=0.2-0.3$,
or much more than the fraction of baryonic matter.

\textbf{9.} As was mentioned in Section 8, weak lensing transforms a circular object into an elliptical image. Statistical analysis of ellipticities and correlations then gives information on the intervening matter field. Early results indicate $\Omega_m\approx0.3$.

\textbf{10.} A direct empirical proof of the existence of dark matter is furnished by weak lensing observations of 1E0657-558, a unique cluster merger. Due to the collision of two clusters, the dissipationless stellar component and the fluid-like X-ray emitting plasma are spatially segregated. The gravitational potential does not trace the plasma distribution which is the dominant baryonic mass component, but rather approximately traces the distribution of galaxies. The center of the total mass is offset from the center of the baryonic mass peaks, proving that the majority of the matter in the system is unseen. In front of the smaller 'bullet' cluster which has traversed the larger one, a bow shock is evident in the X-rays, cf. Figure~2. Two other merging systems with similar characteristics have been seen, although with lower spatial resolution and less clear-cut cluster geometry.\\

   \begin{figure}[htbp]
\includegraphics{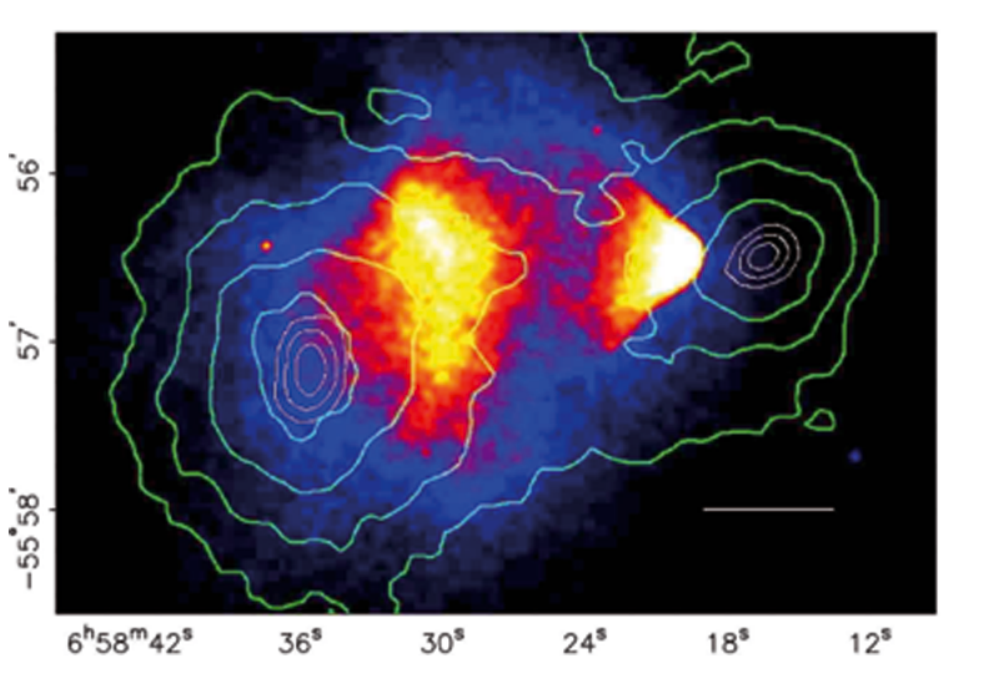}
   \caption{The merging cluster 1E0657-558. On the right is the smaller 'bullet' cluster which has traversed the larger cluster. The colors indicate the X-ray temperature of the plasma: blue is coolest and white is hottest. The green contours are the weak lensing reconstruction of the gravitational potential of the cluster. From D. Clowe \& al., arXiv: astro-ph/0608407, and ApJL 2006.}
      \end{figure}

If only a few per cent of the total mass of the Universe is accounted for by stars and hydrogen clouds, could baryonic matter in other forms make up DM\,? The answer given by nucleosynthesis is a qualified no: all baryonic DM already makes up $\Omega_b$. If there exist particles which were very slow at time $t_{\rm eq}$ when galaxy formation started, they could be candidates for cold dark matter (CDM). If they are \textit{Weakly Interacting Massive Particles} (WIMPs), they became
non-relativistic much earlier than the leptons and decoupled from the hot plasma. For instance, supersymmetry models (SUSY) contain a very large number of particles, of which the lightest
SUSY particle, the \textit{neutralino}, would be stable or perhaps decay slowly into a \emph{gravitino} or an \emph{axino}, or be detectable by its annihilation radiation. \emph{Axions} and heavy neutrinos have also been proposed. A signal worth looking for would be WIMP annihilation radiation from the Galactic disk, from the halo, or perhaps from neighboring dwarf galaxies which exhibit very high ratios of dark/baryonic matter.

Whenever laboratory searches discover a new particle, it must pass several tests in order to be considered a viable DM candidate: it must be neutral, compatible with constraints on self-interactions, consistent with Big Bang nucleosynthesis, and match the appropriate relic density. It must be consistent with direct DM searches and gamma-ray constraints, it must leave stellar evolution unchanged, and be compatible with other astrophysical bounds.

The CDM model has various problems which may require more than a single species of dark matter. In simulations with a single species of CDM the formation of a large number of satellite galaxies is predicted, about 4-10 times more than observed. There is also a problem what density profile describes clusters of galaxies well, "cuspy" or "cored".

\section{Dark Energy}

Hubble's determinations of $H_0$ from observations of the luminosity distances $d_{\rm{L}}(z)\approx z/H_0$ of spiral galaxies at small redshifts proved that the Universe is expanding. At cosmological distances one uses supernovae as standard candles; $d_{\rm{L}}(z)$ must then be redefined so that $H_0$ is replaced by the function $H(a)$ from Equation~(19) which can be fitted to observations. At large redshifts the supernovae appear dimmed, and the parameter $\Omega_{\Lambda}$ comes out large, of the order of 0.7. This agrees with the CMBR results, that $\Omega_{\Lambda}+\Omega_m=1$ with $\Omega_m\approx0.24$.

Moreover, the supernova data indicate that the expansion is not constant, rather the Universe has started to accelerate lately, since $z\approx0.5$.
The second FL Equation~(17) shows, that the expansion can accelerate only if there is a positive term on its right-hand-side: a repulsive cosmological constant $\Lambda$, or something to that effect.

But a cosmological constant is difficult to understand theoretically. The values of $\Lambda$ and $H^2$ are of the same magnitude, or $\rho_{\Lambda}=\Lambda/8\pi G\approx 10^{-47}\,{\rm GeV}^{\,4}$. If $\rho_{\Lambda}$ were even slightly larger, the repulsive force would have caused the Universe to expand so fast that there would not have been enough time for the formation of galaxies. No quantity in physics this small has ever been known before. It is extremely
uncomfortable that $\Lambda$ has to be fine-tuned to a value which differs from zero only in the 52\,nd decimal place.

A second question is why the sum $\Omega_{\Lambda}+\Omega_m$
is precisely 1.0 today when we are there to observe it, after an expansion of some 13~billion years when it was always greater than~1.0. $\Omega_m$ evolves like $a^{-3}$ while $\Omega_{\Lambda}$ remains constant, so why has $\Lambda$ been fine-tuned to come to dominate the sum only now? This is referred to as the \textit{coincidence problem}. One answer is the \emph{Anthropic Principle}: the Universe is as it is because otherwise we would not be here to observe it. But this is not a very satisfactory answer because it makes no testable predictions.

Thirdly, we do not have the slightest idea what the $\Lambda$~energy is, only that it distorts the geometry of the Universe as if it were matter with negative pressure, and it acts as an anti-gravitational force which is unclustered on all scales. We don't know whether $\Lambda$ is constant in time, nor whether its equation of state parameter always is $w_\Lambda=-1$. Actually $w_\Lambda$ is degenerate with $\Omega_{dm}$, we cannot measure them independently.

If $\Lambda$ corresponds to the vacuum energy density $\rho_{vac}$, one should be able to calculate it. In quantum mechanics $\rho_{vac}$ cannot be exactly zero, as is verified experimentally from the quantum noise found in dissipative systems, e.g. in devices based on Josephson junctions. Theoretically, a calculation of $\rho_{vac}$ in quantum field theory is infinite, but the theory is probably valid only up to some scale, e.g. to the
Planck scale, so that $\rho_{vac}\approx10^{74}\,{\rm GeV}^{\,4}$, or at least to the QCD scale  $\rho_{vac}\approx 10^{-3}\,{\rm GeV}^{\,4}$. Both scales disagree catastrophically with the observational value $\rho_{\Lambda}\approx 10^{-47}\,{\rm GeV}^{\,4}$. Why does vacuum energy gravitate so weakly -- a question which is also not removed if $\Lambda$ were exactly zero.

This situation has spurred interest in SUSY models in which cancellations cause $\rho_{vac}= \Lambda=0$ without fine-tuning. But matter as we know it is certainly not supersymmetric and no SUSY particles have been found (yet), so SUSY would then have to be strongly broken. Another approach is to generalize the four-dimensional spacetime so that all particles would be strings in a higher-dimensional manifold, and $\Lambda g_{\mu\nu}$ would be nearly canceled. This might also explain inflation, but unfortunately it can be done in more than $10^{100}$ ways.

If the accelerated expansion is not caused by a cosmological constant, most other explanations  fall in three categories: \textbf{1 --} new \emph{dynamical} dark energy introduced in the energy-momentum tensor $T_{\mu\nu}$ on the right-hand side of Einstein's equation (13); \textbf{2 --} \emph{geometrical} modifications of the spacetime geometry on the left-hand side of (13); \textbf{3 --} new \emph{spatial} conditions taking into account that the Universe is inhomogeneous, and that the apparent expansion appears different in local underdensities than in surrounding superclusters. All present models are rather indistinguishable from acceleration due to a cosmological constant. Observational data confirm that $w\approx-1$, but with large errors.

\textbf{1.} The dynamical explanation could be an all-permeating fluid, a slowly evolving scalar field $\varphi(t)$ with negative pressure, minimally coupled to gravity, having a potential $V(\varphi)$ with an energy density behaving like a decaying cosmological constant, $\Lambda(t)$. \emph{Quintessence} models avoid fine-tuning at Planck time by assuming a slowly rolling scalar field which converges towards a common evolutionary tracker field, almost regardless of the initial conditions. As long as $\varphi(t)$ stays on the tracker solution and is smaller than the background component, $w_{\varphi}$ automatically decreases to a negative value at time $t_{\rm eq}$. The condition for accelerated expansion at present is $w_{\varphi}<-\frac13$, but eventually $w_{\varphi}$ approaches $w_{\Lambda}=-1$. This requires a judicious choice of the potential $V(\varphi)$ with fine-tuned parameters, and then the coincidence problem is still not evaded. Exotic candidates of dark energy may be neutrino condensates and massive vector particles minimally coupled to gravity.

\textbf{2.} General Relativity has been well tested in the Solar system and in        micro-gravity laboratory experiments, so modifications at large scales must recover GR at short distances, and must be able to describe all epochs with mathematically stable solutions: inflation, radiation-domination, matter-domination, and the present acceleration-domination.
Einstein's tensor contains a scalar $R$, the \emph{Ricci scalar}, which might be replaced by some scalar function $f(R)$ meeting the above requirements.

If we assume that we live on a four-dimensional \emph{brane} in a higher-dimensional \emph{bulk} spacetime, new exotic possibilities open up. Particle physics and our observable world would be confined to the brane, but gravitation might move in the bulk. This offers the possibility, that gravitation weakens with time or on large scale, `leaking out' from the brane, and thus causing the observed acceleration.

Special relativity in four-dimensional spacetime forbids faster-than-light particles, \emph{tachyons}, but in higher dimensions nothing hinders tachyons from travelling in the bulk and transporting energy. The potential of the tachyon fields can be chosen to show negative pressure to cause accelerated expansion. There are models in which the field exhibits \emph{phantom} behavior, $w_{\varphi}<-1$, for some time.
In all cases the theory is required to contain no \emph{ghosts}, fields with negative kinetic energy, and the effective EOS should approach $w_{\rm eff}=-1$ in the future.

\textbf{3.} Various models have been proposed how to account for the effect of voids on cosmological scale. Suppose that we are located at or near the center of an underdense region (really an antropocentric assumption which also implies fine-tuning). If the
Universe was homogeneous until structure formation began and small-scale inhomogeneities commenced, the cosmological parameters would develop different values in the void than in surrounding denser clusters. The curvature and the clock rates would be different so that our void would appear to expand faster than the surrounding "walls". Another approach is to try to work out the effect of voids on our observables, whether we are in a void or not.

A third approach considers how to estimate correctly cosmic averages from local measurements. Usually we first choose a metric (eg. RW) for a uniformly averaged universe with an energy-momentum tensor $\langle T_{\mu\nu}\rangle$. In a second step we then calculate the Einstein tensor $G_{\mu\nu}$ with this metric, and substitute $\langle T_{\mu\nu}\rangle$. However, this procedure is wrong, because the first and second steps do not commute. We should rather start from a fine-scale metric which is locally true, then
choose a fine-scale energy-momentum tensor $T_{\mu\nu}$, and finally calculate the "real" $G_{\mu\nu}$ and the averaged $\langle G_{\mu\nu}\rangle$.

Whether such a \emph{back-reaction} from nonlinearities is enough to explain the dark energy effect appears unproven.

\section{Conclusions}

Cosmology is an area in spectacular development, both observationally and theoretically, facing enormous enigmas. How did it all start at the Big Bang, what drove the inflationary expansion, how did inflation stop? What constitutes the dominating gravitating matter, dark matter, and what dark energy effect is causing the present accelerated expansion? Will it halt, or will it dilute the Universe forever? Why are dark matter and dark energy today of the same order of magnitude? Do we live in a four-dimensional spacetime described by General Relativity, or is our spacetime only a brane in a higher-dimensional bulk universe? Cosmology certainly presents the greatest intellectual challenges in our time.

\subsection*{Acknowledgements}

It is a pleasure to acknowledge the expert proofreading help of Prof. Rolf Stabell at the University of Oslo, as well as the hospitality of the Department of Physical Sciences at the University of Helsinki which is providing me with working space and utilities.


\section*{References}
\begin{thebibliography}{10}
\bibitem[1]{Harrison} Harrison E. (1987). Darkness at night. Harvard University Press, Cambridge, MA.
\bibitem[2]{Peacock} Peacock J.A. (1999). Cosmological physics. Cambridge University Press, Cambridge.
\bibitem[3]{Roos} Roos M. (2003). Introduction to Cosmology, Third Edition. John Wiley \& Sons, Ltd, England.
\end{thebibliography}
\end{document}